\documentclass[graphicx]{aastex}

\usepackage{ifthen}
\newboolean{thesis}
\setboolean{thesis}{false} 

\usepackage{amsmath}
\usepackage{natbib} 
\usepackage{color} 
\usepackage{array}
\usepackage{booktabs}
\usepackage{graphicx}
\usepackage{multirow}
\usepackage{rotating}
\usepackage{framed} 
\usepackage{tabularx}
\usepackage{longtable}
\usepackage{booktabs} 
\usepackage{ctable}
\usepackage{subfigure}
\usepackage{rotating}

\ifthenelse {\boolean{thesis}}
{
\usepackage{nonfloat}
}

 \newboolean{outline}
 \newboolean{notes}
 \setboolean{outline}{false} 
 \setboolean{notes}{false}
 \def\2F1{\mbox{$_2${F}$_1$}}

 \newcommand{\BE}{\begin{equation} }
 \newcommand{\BEn}{\begin{enumerate}}
 \newcommand{\BI}{\begin{itemize}}
 \newcommand{\BEA}{\begin{eqnarray}}

 \newcommand{\EE}{\end{equation} } 
 \newcommand{\EEA}{\end{eqnarray}}
 \newcommand{\EEn}{\end{enumerate}}
 \newcommand{\EI}{\end{itemize}}

 \newcommand{\eqn}[1]{Eqn.~\ref{#1}}
  
 \newcommand{\eqns}[2]{Eqns.~\ref{#1}--\ref{#2}}
 \newcommand{\eqnsand}[2]{Eqns.~\ref{#1} and \ref{#2}}

 \newcommand{\fig}[1]{Fig.~\ref{#1}} 
 
 \newcommand{\figs}[2]{Figs.~\ref{#1}--\ref{#2}}

 \newcommand{\gsim}{\mathrel{\hbox{\rlap{\lower.55ex\hbox{$\sim$}} \kern-.3em \raise.4ex \hbox{$>$}}}}

 \newcommand{\lsim}{\mathrel{\hbox{\rlap{\lower.55ex\hbox{$\sim$}} \kern-.3em \raise.4ex \hbox{$<$}}}}

 \newcommand{\nframe}[1]{\ifthenelse{\boolean{outline}}{{{\color{red}\begin{framed}{\begin{enumerate} #1\end{enumerate}}\end{framed}}}}}{}
 
 \newcommand{\note}[1]{\ifthenelse{\boolean{notes}}{\textcolor{red}{(#1)}}\ }{}

 \newcommand{\sect}[1]{Sec.~\ref{#1}}
 \newcommand{\sects}[2]{{Secs.~\ref{#1}--\ref{#2}}}

\begin{document}

\title{Slightly Two or Three Dimensional Self-Similar Solutions}
\author{Re'em Sari, Nate Bode, Almog Yalinewich, Andrew MacFadyen}

\begin{abstract}
Self similarity allows for analytic or semi-analytic solutions to many hydrodynamics problems. Most of these solutions are one dimensional. Using linear perturbation theory, expanded around such a one-dimensional solution, we find self-similar hydrodynamic solutions that are two- or three-dimensional. Since the deviation from a one-dimensional solution is small, we call these slightly two-dimensional and slightly three-dimensional self-similar solutions, respectively. As an example, we treat strong spherical explosions of the second type. A strong explosion propagates into an ideal gas with negligible temperature and density profile of the form $\rho(r,\theta,\phi)=r^{-\omega}[1+\sigma F(\theta,\phi)]$, where $\omega>3$ and $\sigma \ll 1$.  Analytical solutions are obtained by expanding the arbitrary function $F(\theta,\phi)$ in spherical harmonics. We compare our results with two dimensional numerical simulations, and find
good agreement.
\end{abstract}

\keywords{hydrodynamics, shock waves, instabilities}

\section{Introduction}

Astrophysics supplies ample examples of hydrodynamic problems that admit self-similar solutions. In supernovae explosions \citep{KoM90,Che76} a shock wave is created by the release of an immense amount of energy during a short time in the center of an exploding star. When the shock wave propagates into the surrounding medium, the hydrodynamics is described by the the Sedov-Taylor solutions \citep{Sed46,VoN47,Tay50,WaS93}. Gamma-ray bursts provide a relativistic analog of that \citep{BaM76,BeS00,Sar06,PaS06}. If the external medium is spherical, these are one-dimensional solutions. However, if the external density has angular dependence, it will cause the shape of the shock, and the flow behind it, to deviate from sphericity.

An inherently two-dimensional version of this problem is the explosion in half space. Here, space is assumed to be empty on one side of a plane, while the other side is filled with an ideal gas with constant density. A large amount of energy is then released at a point on the surface. This describes the propagation of shockwaves in the process of cratering caused by large impacts on a planetary surface. Qualitatively, this problem and its self-similar nature was described by \cite{ZeR67}, but a two-dimensional self-similar solution was not developed there.

Here, we obtain two-dimensional and three-dimensional self-similar solutions that deviate only slightly from some known one-dimensional solution. We show that when treating such solutions as perturbations, the analysis is analogous to the treatment of stability \citep{RyV87,Goo90,Che90,SWS00,KWS05}. We call these solutions slightly two-dimensional or slightly three-dimensional self-similar solutions. As a working example, we analyze small deviations from sphericity in the case of the strong explosion problem with external density falling as a power law of distance $\rho \propto r^{- \omega}$, where $\omega>3$. Solutions with these values of $\omega$ are known to be self-similar solutions of the second type \citep{WaS93}.

In \sect{sec:6:onedsolution} we briefly review the main features of the one-dimensional solution which serves as the unperturbed solution for our analysis. In \sect{sec:6:multidsolutions}, we discuss the perturbation formalism for this problem and find the slightly two-dimensional and slightly three-dimensional self-similar solutions for density perturbations proportional to a spherical harmonic. These solutions are then demonstrated using the values of $\omega$ and $\gamma$ for which the unperturbed solution is analytic in \sect{sec:6:results}. The solution is then given for small deviations from sphericity with arbitrary angular dependence in \sect{sec:6:arbitrarydensity}, while in \sect{sec:6:shortwavelengthlimit} the small $l$ limit is investigated. Our semi-analytic solutions are then favorably compared with full fluid-dynamic simulations in \sect{sec:6:numericalcomparison}. Finally, in \sect{sec:6:discussion} we give our concluding remarks.

\section{The One-Dimensional Self-Similar Solution}
\label{sec:6:onedsolution}

Here we summarize the formalism leading to the one-dimensional self-similar solution \citep{WaS93}. The discussion here follows \cite{SWS00}. Consider the Strong Explosion Problem in which a large amount of energy is released at the center of a sphere of ideal gas with a density profile decreasing with the distance from the origin according to $\rho=Kr^{-\omega}$, forming a strong outgoing shock wave.

This problem was first investigated by Sedov (1946), Von-Neumann (1947), and Taylor (1950), who found the solutions for $\omega <5$, known as the Sedov-Taylor solutions. \cite{WaS93} showed that the Sedov-Taylor solutions are valid only for $\omega<3$, where the solutions are known as self-similar solutions of Type-I, and contain decelerating shock waves. New, Type-II, self-similar solutions for almost all the range $\omega >3$ containing accelerating shock waves were constructed.

Here we briefly summarize the Type-II solutions for $\omega > 3$. The hydrodynamic equations for an ideal gas with adiabatic index $\gamma$ in spherical symmetry are given by:
\begin{equation}
\begin{array}{c}
(\partial _{t}+u\partial _{r})\rho +\rho r^{-2}\partial _{r}(r^{2}u)=0 \ ,\\
\noalign{\bigskip}\rho (\partial _{t}+u\partial _{r})u+\partial _{r}(\rho
c^{2}/\gamma )=0 \ ,\\
\noalign{\bigskip}(\partial _{t}+u\partial _{r})(c^{2}\rho ^{1-\gamma
}/\gamma )=0 \ ,
\end{array}
\label{hydro}
\end{equation}
where the dependent variables $u$, $c$, and $\rho$ are the fluid velocity, sound velocity, and density, respectively. We now seek a self-similar solution to the hydrodynamic equations (\eqn{hydro}) of the form:
\begin{equation}
\begin{array}{c}
u(r,t)=\dot{R}\xi U(\xi )\ ,\ c(r,t)=\dot{R}\xi C(\xi )\ , \\
\noalign{\bigskip}\rho (r,t)=BR^{\epsilon }G(\xi )\ ,\ p(r,t)=BR^{\epsilon }%
\dot{R}^{2}P(\xi )\ ,
\end{array}
\label{similarityform}
\end{equation}
where $\xi =r/R(t)$ is the dimensionless spatial coordinate, and the length scale $R(t)$ (frequently abbreviated as simply $R$) is the shock radius and satisfies \citep{ZeR67, WaS93}
\begin{equation}
\frac{\ddot{R}R}{\dot{R}^{2}}=\delta\ \  \Rightarrow \ \ \dot{R} \propto R^{\delta }\ ,
\label{Requation}
\end{equation}
where $\delta$ is a constant. The quantities G, C, U, and P, which are defined by Eqns.~\ref{similarityform}, give the spatial dependence of the hydrodynamic quantities. The diverging (exploding) solutions of \eqn{Requation} are
\begin{equation}
R(t)=
\begin{cases}
A(t-t_0)^\alpha, & \text{$\delta<1$} \\ 
\noalign{\bigskip}
Ae^{t/\tau}, & \text{$\delta=1$} \\ 
\noalign{\bigskip} 
A(t_0-t)^\alpha, & \text{$\delta>1$}
\end{cases}
\label{Rsolution}
\end{equation}
where $\alpha =1/(1-\delta )$.

Solutions with $\delta <1$ diverge in infinite time, and $t_{0}$ represents the time of the point explosion, which is usually taken to be $t_0=0$. For $\delta <0$ the shock wave decelerates and for $0<\delta$ it accelerates. For $\delta >1$ the shock wave accelerates so fast that it diverges in a finite time. In this case, $t_{0}$ represents the time of divergence rather than the explosion time. The transition between finite and infinite divergence occurs at $\delta =1$ where we have exponential time dependance \citep{SiM97}.

Substituting \eqn{similarityform} into the hydrodynamic equations (Eqns. \ref{hydro}) and using \eqn{Requation}, one gets regular differential equations for the similarity quantities $U$, $C$, and $G$ (see for example Landau \& Lifshitz) with two free constants, the similarity parameters $\epsilon $ and $\delta $:
\begin{equation}
\label{ducdxi}
\frac{dU}{d\log\xi}=\frac{\Delta_1(U,C)}{\Delta(U,C)}\ ,\
\frac{dC}{d\log\xi}=\frac{\Delta_2(U,C)}{\Delta(U,C)}
\end{equation}
and an explicit expression for the density G:
\begin{equation}
C^{-2}(1-U)^\lambda G^{\gamma-1+\lambda}\xi^{3\lambda-2}={\rm const} \ .
\end{equation}
The functions $\Delta$, $\Delta_1$, and $\Delta_2$ are given by:
\begin{equation}
\begin{array}{c}
\Delta=C^2-(1-U)^2 \ ,\\
\noalign{\bigskip}
\Delta_1=U(1-U)(1-U-\delta)-3UC^2-3C^2(\epsilon+2\delta)/\gamma \ ,\\
\noalign{\bigskip}
\Delta_2=C(1-U)(1-U-\delta)-(\gamma-1)CU(1-U+\delta/2)-\\
\noalign{\medskip}
\phantom{\Delta_2=C(1-U)}\  -C^3+\frac{\textstyle 2\delta-(\gamma-1)\epsilon}{\textstyle 2\gamma}
\frac{\textstyle C^3}{\textstyle 1-U} \ ,
\end{array}
\end{equation}
and the parameter $\lambda$ is
\begin{equation}\label{eqn:lambda}
\lambda=\frac{2\delta-(\gamma-1)\epsilon}{3+\epsilon} \ .
\end{equation}

The similarity parameter $\epsilon$ can be found from the boundary conditions at the strong shock, the Hugoniot jump conditions \citep{LaL87}. Applying these relations to a strong shock one gets $\epsilon =-\omega $, and also
\begin{equation}
U(1)=\frac{2}{\gamma +1}\ ,\ C(1)=\frac{\sqrt{2\gamma (\gamma -1)}}{\gamma +1%
}\ ,\ G(1)=\frac{\gamma +1}{\gamma -1}\ .
\end{equation}
The boundary conditions on the shock do not state any limits on the possible values of the similarity parameter $\delta $. In order to determine the value of this parameter one should distinguish two kinds of similarity flows: Type-I and Type-II, defined first by Zel'dovich \citep{ZeR67}. A solution of Type-I describes the flow in all space and therefore conservation laws must be obeyed by the self-similar solution. One can then deduce $\delta =(\omega -3)/2$, which gives the well-known Sedov-Taylor solutions. However, for $\omega >3$ it is easy to see that the solution obtained with this value of $\delta $ contains an infinite amount of energy and therefore can not describe the flow over the whole space. Therefore, the flow must be Type-II.

In Type-II solutions, there is a region, whose scale relative to the flow characteristic length $R(t)$ goes to zero with time, in which the similarity solution does not describe the physical system. Therefore, for this kind of solution the energy does not have to be conserved in the self-similar solution since this solution does not describe the whole flow. In order that the region which is not self-similar (located around the origin)  does not influence the self-similar solution, the solution must pass through the singular point defined by \citep{ZeR67, WaS93}:
\begin{equation}
U+C=1\ .
\end{equation}

From this singular point requirement, the dependence of $\delta$ upon the parameters $\omega $ and $\gamma $ can be found. It was found \citep{WaS93} that for $\omega>\omega _{g}(\gamma )>3$ there is a value of $\delta $ for which the solution passes through a singular point, and therefore a second type self-similar solution exists.

A fully analytic solution to \eqns{ducdxi}{eqn:lambda} exists for the case where
$\omega=\omega_a(\gamma) \equiv 2(4\gamma-1)/(\gamma+1)$:
\begin{equation}
\begin{matrix}
C(\xi)=\frac{\textstyle\sqrt{2\gamma(\gamma-1)}}{\textstyle\gamma+1}\xi^3
\ ,\ U(\xi)=\frac{\textstyle 2}{\textstyle\gamma+1}\ ,\cr
\noalign{\bigskip}
 G(\xi)=\frac{\textstyle\gamma+1}{\textstyle\gamma-1}\xi^{-8}
\ ,\ P(\xi)=\frac{\textstyle 2}{\textstyle\gamma+1}\ .\cr
\end{matrix}
\end{equation}
For this analytical case the parameter $\delta$ is given by
$\delta=(\gamma-1)/(\gamma+1)$.

\section{Slightly Two- and Three-Dimensional Self-Similar Solutions}
\label{sec:6:multidsolutions}

We now consider small deviations from the spherically symmetric problem discussed in \sect{sec:6:onedsolution}. In general, we wish to solve the problem for an external density perturbation of arbitrary angular dependence, which we parameterize by
\BE
\rho(r,\theta,\phi)=r^{-\omega}[1+\sigma F(\theta,\phi)] \,, \label{eqn:6:rho}
\EE
where $\sigma \ll 1$, $F$ is an arbitrary function of $\theta$ and $\phi$, and $\omega>3$. 

However, before embarking on this general problem we assume in this and the following section that $F(\theta,\phi)=Y_{l,m}$. In \sect{sec:6:arbitrarydensity} these solutions will be used to construct the solution for an arbitrary $F(\theta,\phi)$.

We shall use here the Eulerian perturbation approach. We define the perturbed quantities as the difference between the perturbed solution (i.e., the slightly two-dimensional self-similar solution) and the unperturbed one-dimensional solution at the same spatial point. The derivation of the perturbation equation is similar to the one given by \cite{RyV87}, \cite{Che90}, and \cite{SWS00}. The perturbed hydrodynamic quantities are defined as
\begin{equation}
\begin{array}{c}
\delta \vec{v}(r,\theta ,\phi ,t)=\vec{v}(r,\theta ,\phi ,t)-v_{0}(r,t)\hat{r%
} \ ,\\
\noalign{\bigskip}\delta \rho (r,\theta ,\phi ,t)=\rho (r,\theta ,\phi
,t)-\rho _{0}(r,t) \ ,\\
\noalign{\bigskip}\delta p(r,\theta ,\phi ,t)=p(r,\theta ,\phi ,t)-p_{0}(r,t) \ ,
\end{array}
\end{equation}
where $\vec{v}$, $p$, and $\rho $ are the velocity, pressure, and density in the perturbed solution, while $v_{0}\hat{r}$, $p_0$, and $\rho _{0}$ are the same quantities as in the unperturbed solution. 

We consider perturbations that can be written in a separation of variables form \citep{Cox80}:
\begin{equation}
\begin{array}{c}
\delta \vec{v}(r,\theta ,\phi ,t)=\xi \dot{R}\left[ \delta U_{r}(\xi
)Y_{lm}(\theta,\phi )\hat{r} + \delta U_{T}(\xi ) \nabla_{T}Y_{lm}(\theta,\phi )\right] f \ ,\\
\noalign{\bigskip}\delta \rho (r,\theta ,\phi ,t)=BR^{\epsilon }\delta G(\xi
)Y_{lm}(\theta ,\phi )f \ ,\\
\noalign{\bigskip}\delta p(r,\theta ,\phi ,t)=BR^{\epsilon }\dot{R}%
^{2}\delta P(\xi )Y_{lm}(\theta ,\phi )f \ ,
\end{array}
\label{perturbationsimdef}
\end{equation}
where
\begin{equation}
\nabla _{T}\equiv\hat{\theta}\frac{\partial }{\partial \theta }+\hat{\phi}\frac{1%
}{\sin \theta }\,\frac{\partial }{\partial \phi }
\end{equation}
are the tangential components of the gradient and $R(t)$ is the unperturbed shock radius which still satisfies \eqn{Requation}. The perturbed shock radius, $R(t,\theta ,\phi )$, is given by
\begin{equation}
R(t,\theta ,\phi )-R(t)\equiv \delta R(t,\theta ,\phi)=Y_{l,m}(\theta ,\phi
)R(t)f\ .  \label{defdr}
\end{equation}
\eqnsand{perturbationsimdef}{defdr} define the quantities $\delta U_{r}$, $\delta U_{T}$, $\delta P$, $\delta G$, and $f$. The quantity $f$ measures the fractional amplitude of the perturbation to the shock wave radius. Here we deviate from the standard treatment of stability. There, $f$ is a function of time: if the function $f$ increases with time then the solution is unstable, while if $f$ decreases with time then the solution is stable. However, here, since we demand that the perturbed solution be self-similar, $f$ has to be independent of time.

We linearize the hydrodynamic equation around the unperturbed self-similar solution to get a linear set of equations:
\begin{equation}
\label{matrixform}
\begin{array}{c}
MY^\prime=NY
\end{array}
\end{equation}
\vglue-10pt\noindent
where
\[
\begin{array}{l}
 Y=\left( \begin{matrix}\delta G \cr \delta U_R \cr \delta U_T \cr \delta P \end{matrix} \right)
\ ,\\ M\,=\,\left( \begin{matrix}
\xi(U\hbox{--}1)	&	G\xi	&	0	&	0	\cr
0		&	(U\hbox{--}1)\xi^2G &	0	&	1	\cr
0		&	0	&	(U\hbox{--}1)\xi^2G &	0	\cr
\hbox{--}\frac{\textstyle\gamma\xi(U\hbox{--}1)}{\textstyle G} & 0	&	0
&	\frac{\textstyle\xi(U\hbox{--}1)}{\textstyle P}
\end{matrix} \right)\ ,\\
\noalign{\bigskip\bigskip\bigskip\medskip}
N=\left( \begin{matrix}
\omega\hbox{--}3U\hbox{--}\xi U^\prime&\hbox{--}\xi
G^\prime\hbox{--}3G	&
l(l+1)G				&	0			\cr
P^\prime G^{\hbox{--}1}			&	(1\hbox{--}\delta\hbox{--}2U\hbox{--}\xi
U^\prime)G\xi &
0				&	0			\cr
0				&	0			&
(1\hbox{--}\delta\hbox{--}U)G\xi		&\hbox{--}\xi^{\hbox{--}1}		\cr
-\frac{\textstyle \xi\gamma(U\hbox{--}1)G^\prime} {\textstyle G^2} &
-\xi\left({\textstyle  P' \over \textstyle P}-\gamma{ \textstyle G^\prime\over \textstyle G}	 \right)&
0				&
\xi(U\hbox{--}1){\textstyle P^\prime
\over \textstyle P^2}
\end{matrix}\right) \ ,
\end{array}
\]
and $G$, $U$, and $P$ are defined by \eqn{similarityform}.

Unlike the perturbation equations for stability, the equations above do not contain an unknown parameter. They are, in fact, a special case of the equations used in \cite{SWS00}, but with the perturbation growth rate set to $q=0$. In that sense they are similar to the equations of \cite{OrS09} for discretely self-similar solutions. Instead, a new parameter $d=\sigma/f$ appears in the shock boundary conditions (the linearized Hugoniot jump conditions):
\begin{equation}
\begin{array}{c}
\delta G(1)=\frac{\textstyle\gamma +1}{\textstyle\gamma -1}(d-\omega)
-G^{\prime }\quad ,\quad
\delta U_{r}(1)=-U^{\prime } \\
\noalign{\bigskip}
\delta U_{T}(1)=-\frac{\textstyle 2}{\textstyle \gamma +1}
\quad ,\quad
\delta P(1)=\frac{\textstyle 2}{\textstyle\gamma +1}\left( 2+d-\omega \right) -P^{\prime } \ .
\end{array}
\label{boundaryshock}
\end{equation}

For any value of the parameter $d$ one can integrate \eqn{matrixform} beginning at the shock front using the shock boundary conditions. However, the singular point of the unperturbed solution, $\xi_{c}$, where $C+U=1$, is also a singular point of the perturbed solution. Therefore, in general, such integration will diverge at the sonic point $\xi_{c}$. Only for specific values of the parameter $d$, where an additional boundary condition at the singular point is satisfied, is the solution regular. These are the physical values for the parameter $d$.

Technically, solving these equations is easier than the equivalent perturbation case. The reason is that the unknown parameter $d$ appears only in the shock boundary condition, and is absent from the differential equations. We can therefore solve these equations starting from the sonic point outward, and find the three independent solutions that are nonsingular at $\xi_c$. Then we can find a linear combination of these three solutions, and the value of $d$ that can solve the four boundary conditions at the shock.

\section{Results}
\label{sec:6:results}

\begin{figure}
\begin{center}
\includegraphics[width=\columnwidth]{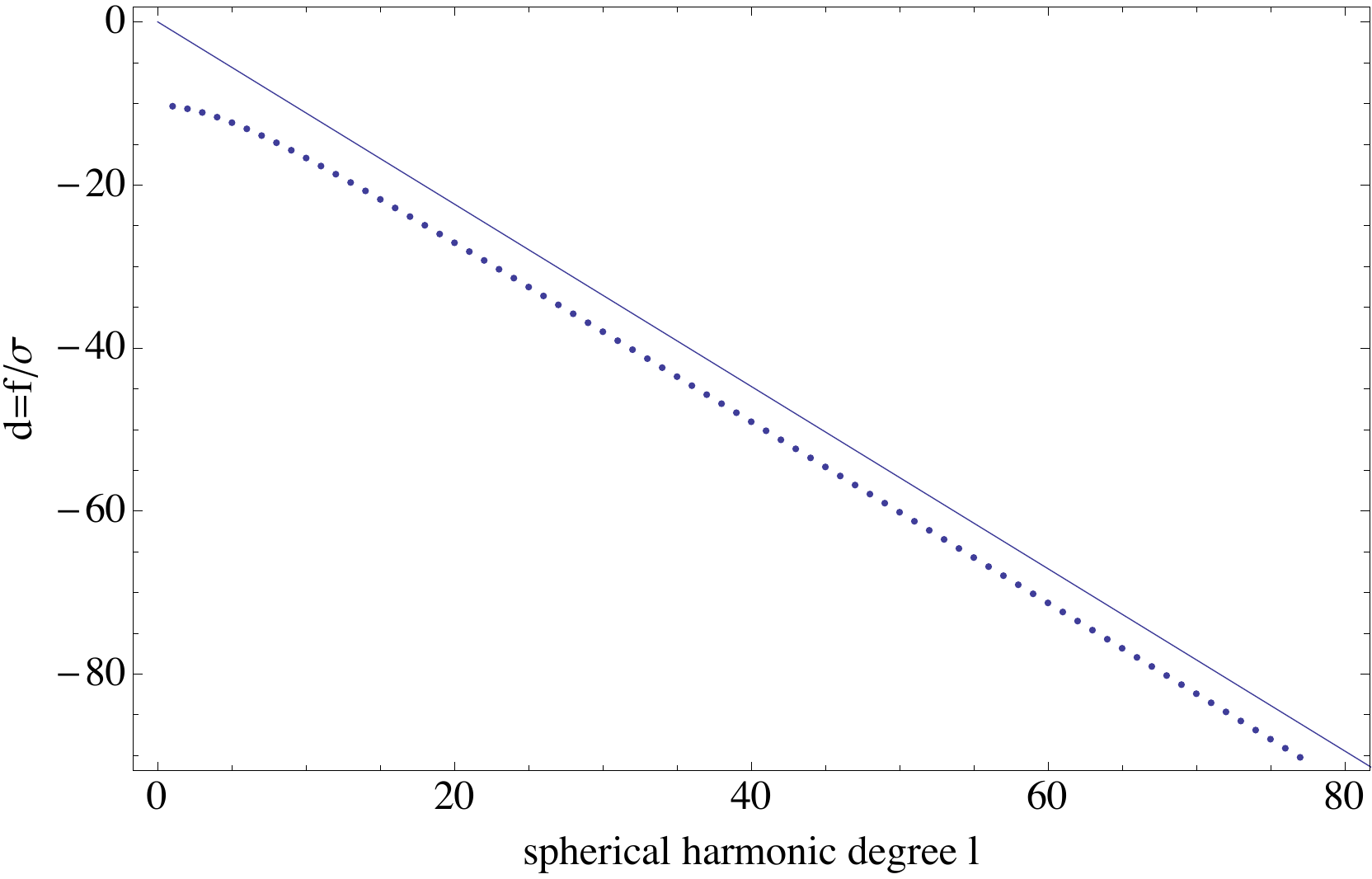}
\caption[Dots show $d$ as a function of $l$]{
Dots show $d$ as function of $l$ as obtained by solving the differential equations for $\gamma=5/3$ and $\omega=17/4$. For small $l$ we have $d \cong -10.$, while for large $l$, i.e., short wavelength, we obtain a linear relation: $d=-\sqrt{5/4}l$ (solid line).
\label{d(l)}
}
\end{center}
\end{figure}

For convenience we investigate the case $\gamma=5/3$, $\omega =17/4$, where the unperturbed solution is analytic. For $l=1$ we obtain $d=-11.2$. This means that the fractional amplitude of perturbations in the shock wave position, $f$, are an order of magnitude smaller than the fractional amplitude of perturbations in the external density $\sigma$. The negative sign implies that at angles where the external density is higher, the shock wave position is retarded. This is expected intuitively. From the shock boundary conditions, we infer that the pressure at these angles is also lower. For $l=2$ we find $d=-11.6$, and for $l=3$ we find $d=-12.1$. A plot of $d$ as function of $l$ is given in \fig{d(l)}.

\section{Extension to Arbitrary Angular Dependence}
\label{sec:6:arbitrarydensity}

The analysis above is limited to external density perturbations whose angular dependence is a spherical harmonic. This is necessary in order to obtain separation between the angular and radial dependencies. However, since we are dealing with linear perturbations, any arbitrary angular dependence can be expanded into a sum of spherical harmonics, each of which could be solved in the method described in the previous section. Then, the solutions can be summed, leading to the perturbation solution for external density perturbations with arbitrary angular dependence.

As an example, we consider the following problem: A strong point-like explosion is launched into a surrounding which has a density on one side of a plane slightly different from the density on the other side of the plane. In our notation this is $\rho \propto r^{-17/4}(1+\sigma H(\theta))$ where $H(\theta)=1$ for $\theta<\pi/2$ and $H(\theta)=-1$ for $\theta>\pi/2$. The point explosion in half space could be thought of as an extreme version of this density profile with $\sigma=1$. However, our solution formalism applies only for slightly two-dimensional cases where $\sigma \ll 1$.

\begin{figure}
\begin{center}
\includegraphics[width=\columnwidth]{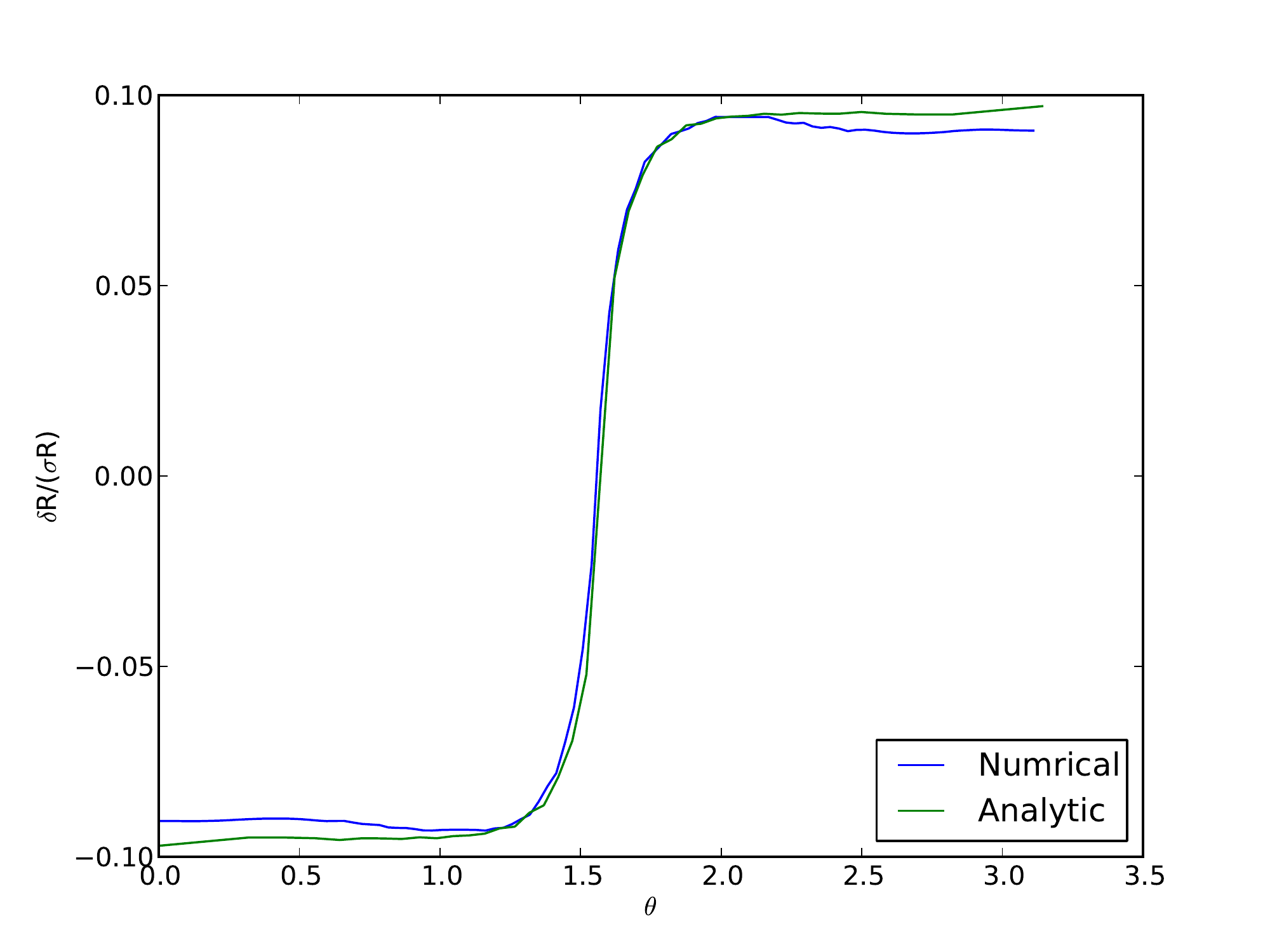}
\caption[The fractional deviation of the shock position as function of $\theta$ for the Heaviside density distribution]{
The analytic (green) and numerical (blue) fractional deviation of the shock position as function of $\theta$ for the Heaviside density distribution (\eqn{eqn:6:heavisidedensity}) in units of $\sigma$. The numerical result is obtained for $\sigma=0.01$, and is discussed in \sect{sec:6:numericalcomparison}, while the analytic solution comes from \eqn{eqn:6:deltaRoR}. It can be seen in both curves that, roughly speaking, the shock is composed of two hemispheres, connected smoothly over a short angular scale of less than $0.1$ radians (FWHM). 
\label{shs}
}
\end{center}
\end{figure}

Such a density profile could be expanded in spherical harmonics as
\begin{equation}
H(\theta)=\sum_{n=0}^\infty {\pi \sqrt{4n+3}  \over \Gamma(1/2-n)\Gamma(2+n)}Y_{2n+1,0}(\theta,0) \ .\label{eqn:6:heavisidedensity}
\end{equation}
The shape of the shock, $R+\delta R(\theta)$, deviates from its unperturbed value $R$ by 
\begin{equation}
{\delta R(\theta) \over R}=\sigma \sum_{n=0}^\infty {\pi \sqrt{4n+3}  \over d(2n+1) \Gamma(1/2-n)\Gamma(2+n)}Y_{2n+1,0}(\theta,0) \ . \label{eqn:6:deltaRoR}
\end{equation}
This shape is plotted in \fig{shs} for both this analytic solution and for the numerical solution discussed in \sect{sec:6:numericalcomparison}. To make the analytic curve the sum was taken from $n=0$ to $n=50$.

\section{Short Wavelength Limit}
\label{sec:6:shortwavelengthlimit}

Because the flow does not vary in the short wavelength limit, we may treat the matrices $M$ and $N$ as constants close to the shock front. By using the unperturbed values of the state variable at the shock, we find the four independent modes of the problem:
\begin{equation}
\lambda=-8, 3, \pm\sqrt{2\gamma\over\gamma+1}l\,.
\end{equation}
The first two are independent of $l$ and indicate that the state functions, close to the shock, vary on the scale $R$, regardless of the wavelength of the perturbation. However, the other two are linear in $l$ meaning that close to the shock the state functions vary over small scales of order $R/l$. Therefore, for large $l$, the positive mode is growing inward very rapidly, and thus can not exist physically. For this reason we demand that the perturbation has no component along this mode on the shock front by requiring it to be written as a linear combination of the three eigenvectors associated with the other modes. This provides the extra boundary condition at the shock that allows us to determine $d$. Performing this calculation we find that for general $\omega$ and $\gamma$
\begin{equation}
\label{eqn:6:dofllargel}
d=-\sqrt{{2\gamma\over\gamma+1}}l
\end{equation}
in the limit $l\gg 1$. We plot the general solution of $d(l)$ in \fig{d(l)} for the case discussed in \sect{sec:6:results}, along with the short wavelength limit described by \eqn{eqn:6:dofllargel}. It can be seen that the agreement is good.

\section{Comparison with 2D Numerical Hydrodynamical Simulations}
\label{sec:6:numericalcomparison}

To compare the analytic solution presented in \sects{sec:6:onedsolution}{sec:6:shortwavelengthlimit} to numerical results we used the PLUTO hydrodynamic code \citep{Mignone+07} to simulate an explosion on a weakly discontinuous surface with $\sigma=0.01$ (see \eqn{eqn:6:rho}). Again for convenience we consider the case where the unperturbed solution is analytic: $\omega=17/4$ and $\gamma=5/3$.

\begin{figure}
\begin{center}
\includegraphics[width=\columnwidth]{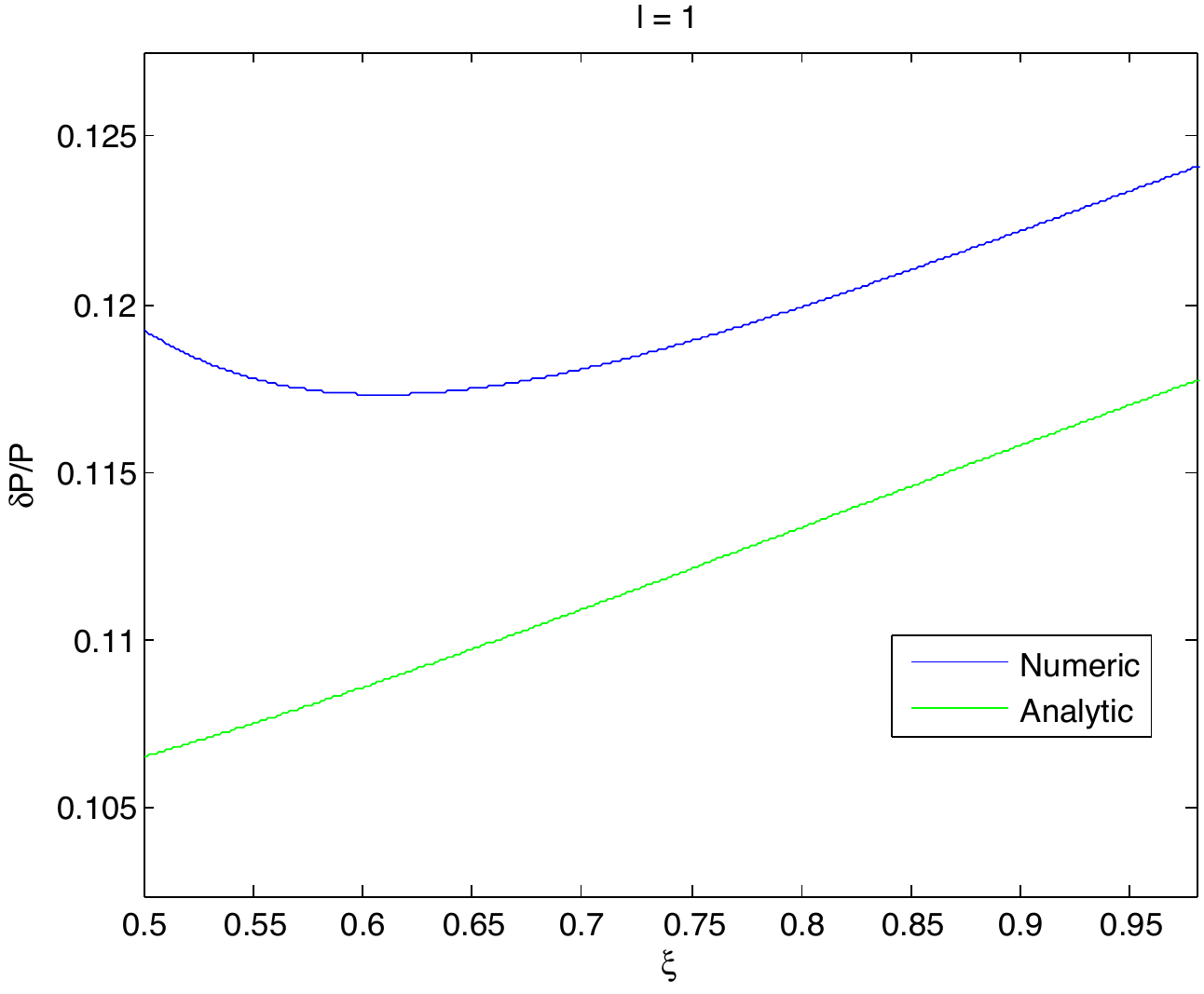}
\caption[Comparison between the numerical (blue) and analytic (green) solutions for the normalized $l=1$ self-similar pressure perturbation as a function of the self-similar variable $\xi$ for a Heaviside initial density distribution.]{Comparison between the numerical (blue) and analytic (green) solutions for the normalized $l=1$ self-similar pressure perturbation as a function of the self-similar variable $\xi$ for a Heaviside initial density distribution (\eqn{eqn:6:heavisidedensity}). Good agreement between the simulation and the analytic solution is found in the region of self-similarity ($0.76=\xi_c\le\xi\le1$). Details of the simulation are discussed in \sect{sec:6:numericalcomparison}, while the analytic solution is described in \sects{sec:6:onedsolution}{sec:6:shortwavelengthlimit}. 

\label{dPoP:Leq1}
}
\end{center}
\end{figure}

The computational mesh had $10^5$ cells in the radial direction, and $100$ cells in the tangential direction. The inner radius was $10^{-6}$ and the outer radius was $1$. The smallest angle was $0$ and the largest $\pi$. The radius of the initial hot spot was $2\times10^{-4}$, and the pressure there was $10^{18}$, whereas outside the hot spot the pressure was $1$. The Riemann solver used was hllc.

The numerical and analytic results are compared in \figs{shs}{dUthetaoU:Leq2}. Though the self-similar solution is valid everywhere, the deeper one looks into the flow, the longer it takes for the physical flow to approach this solution. Therefore, at any finite time, there exists an inner region that is not in agreement with the self-similar solution. Our comparison tends to reflect these points and in all cases there is agreement to within $10\%$ throughout a significant fraction of the flow.

In particular, in \fig{shs} we compare the deviation of the shock radius from the unperturbed solution to the shock radius in units of $\sigma$ in both cases. Analytically this function is independent of $\sigma$ (see \eqn{eqn:6:deltaRoR}). Near the interface of the media ($\theta=\pi/2$) the two solutions match well, while near the poles ($\theta=0,\pi$) the two differ by approximately $10\%$.

\begin{figure}
\begin{center}
\includegraphics[width=\columnwidth]{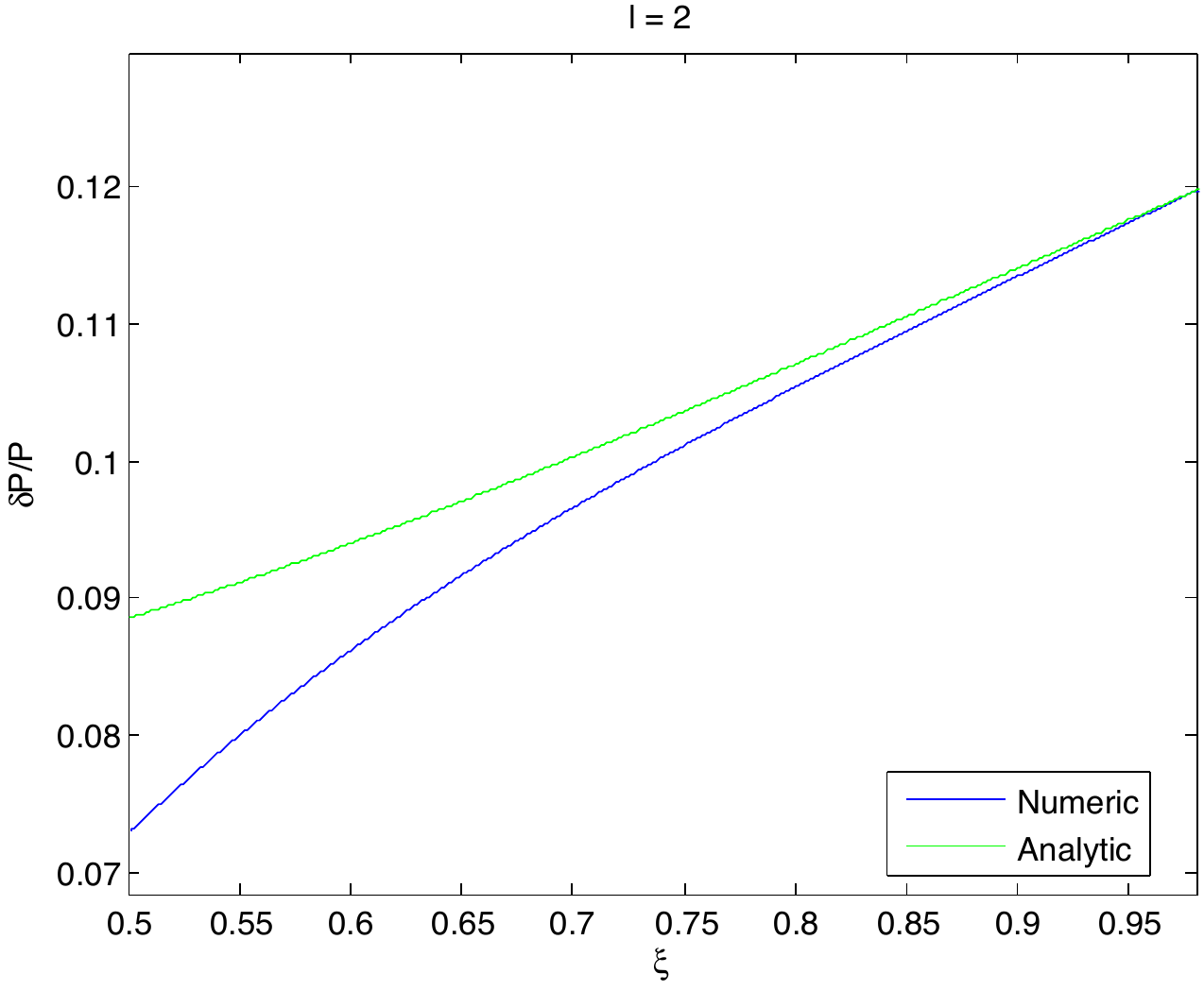}
\caption[Comparison between the numerical (blue) and analytic (green) solutions for the normalized $l=2$ self-similar pressure perturbation as a function of the self-similar variable $\xi$ for a Heaviside initial density distribution.]{Comparison between the numerical (blue) and analytic (green) solutions for the normalized $l=2$ self-similar pressure perturbation as a function of the self-similar variable $\xi$ for a Heaviside initial density distribution (\eqn{eqn:6:heavisidedensity}). Good agreement between the simulation and the analytic solution is found in the region of self-similarity ($0.76=\xi_c\le\xi\le1$).  Details of the simulation are discussed in \sect{sec:6:numericalcomparison}, while the analytic solution is described in \sects{sec:6:onedsolution}{sec:6:shortwavelengthlimit}. 

\label{dPoP:Leq2}
}
\end{center}
\end{figure}

\begin{figure}
\begin{center}
\includegraphics[width=\columnwidth]{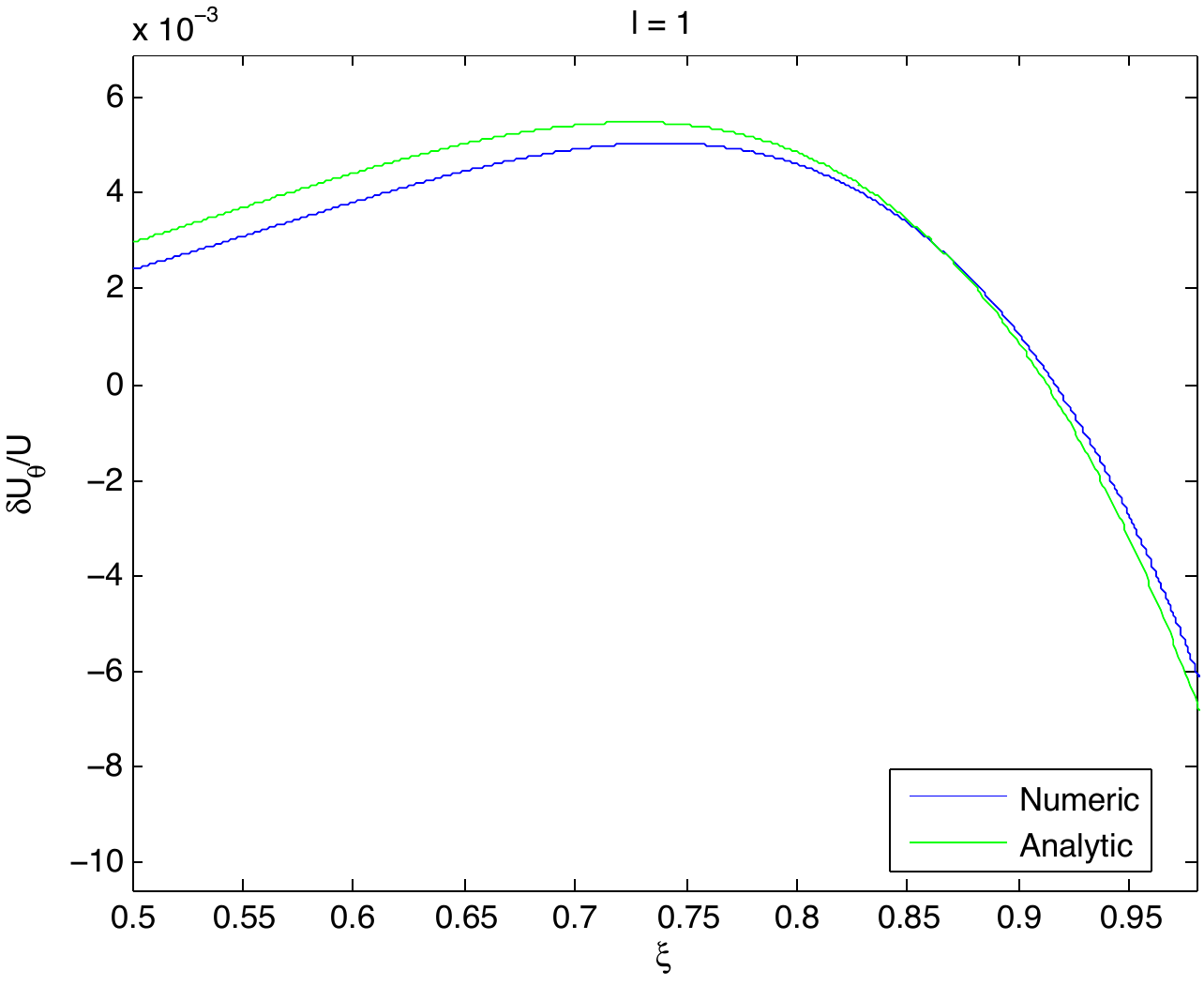}
\caption[Comparison between the numerical (blue) and analytic (green) solutions for the normalized $l=1$ self-similar fractional angular velocity as a function of the self-similar variable $\xi$ for a Heaviside initial density distribution.]{Comparison between the numerical (blue) and analytic (green) solutions for the normalized $l=1$ self-similar fractional angular velocity as a function of the self-similar variable $\xi$ for a Heaviside initial density distribution (\eqn{eqn:6:heavisidedensity}). Good agreement between the simulation and the analytic solution is found even well outside the region of self-similarity ($0.76=\xi_c\le\xi\le1$).  Details of the simulation and its parameters are discussed in \sect{sec:6:numericalcomparison}, while the analytic solution is described in \sects{sec:6:onedsolution}{sec:6:shortwavelengthlimit}. 

\label{dUthetaoU:Leq1}
}
\end{center}
\end{figure}

\begin{figure}
\begin{center}
\includegraphics[width=\columnwidth]{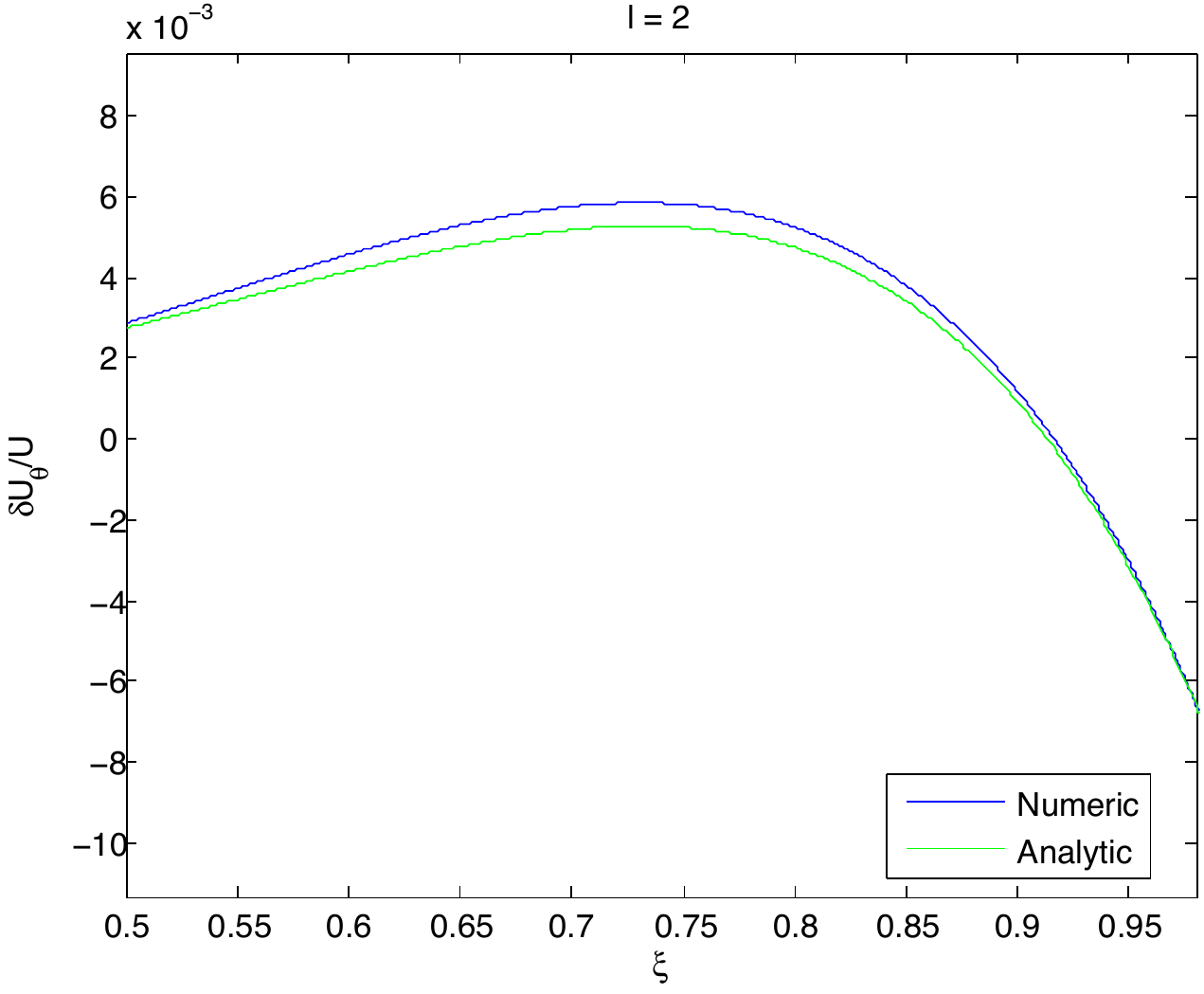}
\caption[The fractional deviation of the shock position as function of $\theta$ for the Heaviside density distribution]{Comparison between the numerical (blue) and analytic (green) solutions for the normalized $l=2$ self-similar fractional angular velocity as a function of the self-similar variable $\xi$ for a Heaviside initial density distribution (\eqn{eqn:6:heavisidedensity}). Good agreement between the simulation and the analytic solution is found even well outside the region of self-similarity ($0.76=\xi_c\le\xi\le1$). Details of the simulation are discussed in \sect{sec:6:numericalcomparison}, while the analytic solution is described in \sects{sec:6:onedsolution}{sec:6:shortwavelengthlimit}.

\label{dUthetaoU:Leq2}
}
\end{center}
\end{figure}

In the following four figures (\fig{dPoP:Leq1}, \fig{dPoP:Leq2}, \fig{dUthetaoU:Leq1}, and \fig{dUthetaoU:Leq2}) we compare the fractional deviation of the pressure and angular velocity for both $l=1$ and $l=2$. As expected, as is the case with the general solution plotted in \fig{shs}, discrepancies between the numerical and analytic work are always less than $10\%$. Descriptions of the specific cases are given in the captions.

\section{Discussion}
\label{sec:6:discussion}

We have considered the problem of a strong shock propagating into a slightly aspherical medium made up of a density with a spherically symmetric radial power-law plus a perturbation of arbitrary angular dependence, and solved for the Type-II self-similar solution. Such an external medium has a density profile $\rho(r,\theta,\phi)=r^{-\omega}[1+\sigma F(\theta,\phi)]$, where $\omega>3$, $\sigma \ll 1$, and $F$ is an arbitrary function of $\theta$ and $\phi$. Because the perturbations are small, the hydrodynamic equations can be linearized around the unperturbed solution. This then allows us to expand $F$ as a series in spherical harmonics, and solve the problem term by term. In this way the general problem is reduced to one which includes only perturbations $F(\theta,\phi)\propto Y_{lm}(\theta,\phi)$.

The linearized self-similar equations are presented for this simpler case, $F(\theta,\phi)\propto Y_{lm}(\theta,\phi)$, along with the appropriate boundary conditions. There is a unique solution to these equations which depends on a single parameter $d$, which is determined by the requirement that the solution would pass smoothly through the sonic point. That the only dependence on $d$ is in the boundary conditions makes these equations particularly straight-forward to solve.

We demonstrate this process on a specific example which deviates from spherical symmetry by a weak step function in the outside density across a plane containing the initial explosion. As expected, instead of the shock being spherical, it is composed of two hemispheres smoothly connected across the plane of the discontinuity. Our 2D hydrodynamical simulations agree well with this solution.

\acknowledgements We thank Yonatan Oren for helpful discussions. This research was partially supported by ERC and IRG grants and by a Packard fellowship. RS is a Guggenheim fellow and a Radcliffe fellow. AIM acknowledges support from NSF grant AST-1009863 and NASA grant NNX10AF62G.

\bibliographystyle{apalike}
\bibliography{similarity.bib}

\end{document}